\documentstyle[12pt]{article}

\newcommand{\be}{\begin{equation}}
\newcommand{\ee}{\end{equation}}
\newcommand{\ba}{\begin{array}}
\newcommand{\ea}{\end{array}}
\newcommand{\bea}{\begin{eqnarray}}
\newcommand{\eea}{\end{eqnarray}}
\newcommand{\none}{\nonumber \\}

\newcommand{\der}{\partial}

\newcommand{\ajo}[2]{\frac{\der #1}{\der #2}}
\newcommand{\ava}[2]{\frac{\delta #1}{\delta #2}}

\newcommand{\pdfi}[1]{\frac{d #1}{d \phi}}

\newcommand{\half}{{1\over2}}

\newcommand{\pihat}{\hat{\Pi}}

\begin{document}

\thispagestyle{empty}
\mbox{}\\[1cm]
\begin{center}
{\bf EXACT QUANTUM STATES FOR ALL TWO-DIMENSIONAL DILATON 
GRAVITY THEORIES}
\end{center}
\vspace{0.5 cm}
\begin{center}
{\sl by\\
}
\vspace*{0.50cm}
{\bf
Domingo Louis-Martinez}

\vspace*{0.50cm}
{\sl
Department of Physics, University of British Columbia\\ 
Vancouver, BC, Canada, V6T 1Z1

{[e-mail: martinez@physics.ubc.ca]}\\[5pt]
}

\end{center}
\bigskip\noindent
{\large
ABSTRACT
}
\par
\noindent
It is shown that the recently obtained quantum wave functionals in 
terms of the CJZ variables
for generic 2d dilaton gravity are equivalent to the previously 
reported exact quantum wave 
functionals in geometrical variables. A third representation of 
these exact quantum states
is also presented.

\newpage

\pagenumbering{arabic}

\section{Introduction}

Two dimensional (2d) theories of gravity have attracted 
a lot of attention 
in recent years as a simplified theoretical model for studying 
Hawking evaporation
of black holes and the information loss problem \cite{0}.
It is believed that these theories could provide some insight
into the problems that appear in the quantization of Einstein's 
theory of gravity \cite{9}.

The exact quantum wave functionals for generic 2d dilaton gravity  
were obtained in \cite{1}. They were given as functionals of the
 spatial
component $g_{11}$ of the metric and the dilaton field $\phi$ as,

\be
\Psi [ C; g_{11}, \phi ] = \chi[C] \exp \left[ {i \over \hbar}
\int_{x_-}^{x_+} dx \left( Q + \phi' \ln\left( {2\phi' - 
Q \over 2\phi' + Q} \right) \right) \right]
\label{eq1}
\ee

\noindent where $C$ is a parameter, $\chi$ is any function of $C$, and,

\be
Q = 2 \sqrt { (\phi')^2 + (C - j (\phi)) g_{11}}
\label{eq2}
\ee

The two dimensional spacetime is assumed to be locally a direct
 product 
$R\times\Sigma$,
where the spatial manifold $\Sigma$ can be either open or closed.
The spatial coordinate $x$ runs from $x_-$ to $x_+$.

Notice that for any 2d dilaton gravity theory the action can
 always be
written as \cite{1}:

\be
S = \int d^2x \sqrt{-g} ( \phi R + V(\phi))
\label{eq3}
\ee

\noindent where $R$ is the scalar curvature.

For spherically symmetric gravity \cite{2}
 $V = \Lambda \phi^{-\half}$, 
for the Jackiw-Teitelboim model \cite{3}
 $V = \Lambda\phi$  
and for the string inspired model \cite{4,5} $V = \Lambda$.
Two-dimensional dilaton gravity models with nonsingular
 black hole 
solutions have also been found \cite{11}.

The functions $V$ and $j$ are related as follows:

\be
\pdfi{j(\phi)} = V(\phi)
\label{eq4}
\ee

A gauge theoretical formulation for string-inspired gravity
 was developed in
\cite{5}. In \cite{6} it was proven that the quantum wave
 functionals 
obtained in \cite{5} are equivalent to the ones of \cite{1} 
(for string
inspired model $j = \Lambda\phi$).

Recently, Barvinsky and Kunstatter \cite{7} generalized the
 work of \cite{9}
and found an exact expression for the quantum wave functionals
 for all 2d dilaton gravity
theories (\ref{eq3}) in the CJZ  \cite{9} variables:

\be
\Psi [C; \eta^0, \eta^1] = \chi[C] \exp \left[ {i \over \hbar}
\int_{x_-}^{x_+} dx \omega (\eta^2) 
\left( \eta^0 (\eta^1)' - \eta^1 (\eta^0)' \right) \right]
\exp \left( {i \over \hbar} \alpha\beta \right)
\label{eq5}
\ee

\noindent where the function $\omega$ is defined by the equation:

\be
C - j \left( \half \eta^2 \omega (\eta^2) \right) =
 {\eta^2 \over 4}
\label{eq6}
\ee

\noindent $\alpha$ is a constant (in \cite{7}, due to
 boundary conditions, 
$\alpha$ was chosen to be 
equal to $j^{-1}(C)$ ) and,

\be
\left. \beta = 2 \tanh^{-1} \left({\eta^0 \over \eta^1} \right) 
\right|_{x_-}^{x_+} 
\ee

Notice that,

\be
\eta^2 \equiv (\eta^1)^2 - (\eta^0)^2
\ee

As can be seen from (\ref{eq1}) or (\ref{eq5}), for all 2d
 dilaton gravity theories the exact solutions
are also the lowest order WKB solutions. 

The purpose of this paper is to provide a proof of the
 equivalence of (\ref{eq1})
and (\ref{eq5}). A third representation for the exact
 quantum states in terms of the dilaton
field and the momentum conjugate to the spatial component
 of the metric is also presented.

\section{Proof of the equivalence}

Assume the classical phase space variables $(q,p)$ and
 $(Q,P)$ are 
related by a canonical transformation generated by
 the function $F(q,Q)$:

\bea
p =  \ajo{F}{q} \hspace{2.0cm}P = - \ajo{F}{Q}
\label{eq71}
\eea

\noindent where,

\be
\frac{\der^2 F}{\der q \der Q} \not= 0
\label{eq7}
\ee

Assume that the eigenvectors of the operators $\hat{q}$ form
a basis and that the eigenvectors of $\hat{Q}$ also form a 
basis
\footnote{For canonical transformations 
of a certain type \cite{8} the transformation
function relating the two bases is given by \cite{8}:
$\exp \left( -{i \over \hbar} F(q,Q) \right)$.}.
A vector $\mid \psi>$ can be defined by its wave function
 $\psi(q)$ in
one basis or by its wave function $\Psi(Q)$ in the other.

Notice that if $\psi(q)$ is determined (up to a
 multiplicative
constant) by the equations:

\be
-i \hbar \ajo{}{q} \psi(q) = p(q) \psi(q)
\label{eq10.1}
\ee

\noindent and $\Psi(Q)$ by the equations:

\be
-i \hbar \ajo{}{Q} \Psi(Q) = P(Q) \Psi(Q)
\label{eq10.2}
\ee

\noindent where $P(Q)$ in (\ref{eq10.2}) is given by,

\be
\left. P(Q) = - \ajo{F}{Q}(q,Q) \right|_{q = q(Q)}
\ee

\noindent and $q = q(Q)$ are the solutions of the equations:

\be
p(q) - \ajo{F}{q}(q,Q) = 0
\label{eq12}
\ee

\noindent then,

\be
\left. \Psi(Q) = \left( e^{-{i \over \hbar} F(q,Q)} \psi(q) 
\right) \right|_{q= q(Q)}
\label{eq11}
\ee

Indeed, it is easy to see that (\ref{eq11}) satisfies
 (\ref{eq10.2}).

The quantum wave functional (\ref{eq1}) is the solution of
 the equations \cite{1}:

\be
-i \hbar {\delta\quad \over \delta\phi} \Psi[C; \rho, \phi] = 
{g[\rho, \phi] \over Q[C; \rho, \phi]} \Psi[C; \rho, \phi]
\label{eq13}
\ee

\be
-i \hbar {\delta\quad \over \delta\rho} \Psi[C; \rho, \phi] =
Q[C; \rho, \phi] \Psi[C; \rho, \phi]
\label{eq14}
\ee

\noindent where,

\be
\rho = \half \ln(g_{11})
\label{eq14.1}
\ee

\noindent and the functions $g$ and $Q$ are,

\be
g[\rho, \phi] = 4\phi'' - 4\phi'\rho' - 2V(\phi)e^{2\rho}
\label{eq15}
\ee

\be
Q[C; \rho, \phi] = 2\sqrt{(\phi')^2 + (C - j(\phi))e^{2\rho}}
\label{eq16}
\ee

The canonical transformation relating the geometrical phase
 space variables\\
$(\rho, \phi, \Pi_\rho, \Pi_\phi)$ and the CJZ variables
 $(\eta^0, \eta^1, P_0, P_1)$
is defined as follows \cite{9}:

\be
e^{2\rho} = P^2 \equiv P_{1}^2 - P_{0}^2
\label{eq17.1}
\ee

\be
\phi' = \half (\eta^0 P_{1} + \eta^1 P_{0})
\label{eq17.2}
\ee

\be
\Pi_{\rho} = - \eta^0 P_{0} - \eta^1 P_{1}
\label{eq17.3}
\ee

\be
\Pi_{\phi} = {2 \over P^2} (P_{0}P'_{1} - P_{1}P'_{0})
\label{eq17.4}
\ee

The generating functional $F[\rho, \phi; \eta^0, \eta^1]$
 can be written as,

\bea
F[\rho, \phi; \eta^0, \eta^1] & = & 
\int_{x_-}^{x_+} dx \left[ 2\sqrt{(\phi')^2 +
 {\eta^2 \over 4}e^{2\rho}}
+ \phi' \ln \left( {\phi' - \sqrt{(\phi')^2 + 
{\eta^2 \over 4}e^{2\rho}} \over 
\phi' + \sqrt{(\phi')^2 + {\eta^2 \over 4}e^{2\rho}}} 
\right) - \right. \none
& & \  \hspace{2.0cm}
- \left. {2 (\phi - \alpha) \over \eta^2} 
\left( \eta^0 (\eta^1)' - \eta^1 (\eta^0)' 
\right) \right] \label{eq18}
\eea

\noindent where $\alpha$ is a constant.

Notice that from (\ref{eq18}) it follows that:

\bea
{\delta F \over \delta\rho} & = & 2\sqrt{(\phi')^2 + 
{\eta^2 \over 4}e^{2\rho}}
\none \\
{\delta F \over \delta\phi} & = & 
{2(\phi'' - \phi'\rho') -\phi' (\ln \eta^2)' 
\over \sqrt{(\phi')^2 + {\eta^2 \over 4}e^{2\rho}}} - 
{2 \over \eta^2}(\eta^0(\eta^1)' - \eta^1(\eta^0)')
\label{eq18.1}
\eea

For the particular system that we are studying 
Eqs (\ref{eq12}) take the form,

\be
Q[C; \rho, \phi] - 
{\delta F \over \delta\rho}(\rho, \phi; \eta^0, \eta^1) = 0
\label{eq19.1}
\ee

\be
{g[\rho, \phi] \over Q[C; \rho, \phi]} - 
{\delta F \over \delta\phi}(\rho, \phi; \eta^0, \eta^1)
= 0
\label{eq19.2}
\ee

From these equations we can find $\phi$ and $\rho$

 as functionals of $\eta^0$ and $\eta^1$.
In particular we find that,

\be
C - j(\phi) = {\eta^2 \over 4}
\label{eq20}
\ee

Consider now the right hand side of Eq (\ref{eq11}).
 From (\ref{eq18}) and (\ref{eq1})
it follows that:

\be
\left. \left( e^{-{i \over \hbar}F[\rho, \phi; \eta^0, \eta^1]}
 \Psi[C; 
\rho, \phi] \right) \right|_{C - j(\phi) ={\eta^2 \over 4}} = 
\exp \left[ {i \over \hbar}
 \int_{x_-}^{x_+} dx {2 (\phi(\eta^2) - \alpha) \over \eta^2} 
(\eta^0 (\eta^1)' - \eta^1 (\eta^0)') \right]
\label{eq21}
\ee

\noindent where, $\phi = \phi(\eta^2)$ is the solution
 of the equation (\ref{eq20}).

Define the function $\omega$ in the following way:

\be
\phi(\eta^2) \equiv \half \eta^2 \omega(\eta^2)
\label{eq22}
\ee

Substituting (\ref{eq22}) into (\ref{eq21}) we finally
 obtain,

\be
\Psi [C; \eta^0, \eta^1] = \chi[C] \exp \left[ {i \over \hbar}
\int_{x_-}^{x_+} dx \omega (\eta^2) 
\left( \eta^0 (\eta^1)' - \eta^1 (\eta^0)' \right) \right]
\exp \left( {i \over \hbar} \alpha\beta \right)
\ee

\noindent where,

\be
C - j \left( \half \eta^2 \omega (\eta^2) \right) = 
{\eta^2 \over 4}
\ee

\noindent and,

\be
\left. \beta \equiv - 2 \int_{x_-}^{x_+} dx {(\eta^0(\eta^1)' - 
\eta^1(\eta^0)') \over \eta^2}
= 2 \tanh^{-1}
 \left( {\eta^0 \over \eta^1} \right) \right|_{x_-}^{x_+}
\label{eq100}
\ee

Therefore, we have obtained (\ref{eq5}) directly
 from (\ref{eq1}).

Viceversa, since the wave functional in terms of the 
CJZ variables (\ref{eq5}) is the
solution of the equations:

\bea
-i \hbar {\delta\quad \over \delta\eta^0} 
\Psi[C; \eta^0, \eta^1] & = &
2 \left( \omega + \eta^2 \ajo{\omega}{\eta^2} \right) 
(\eta^1)' \Psi[C; \eta^0, \eta^1]
\label{eq23.1} \\
-i \hbar {\delta\quad \over \delta\eta^1}
 \Psi[C; \eta^0, \eta^1] & = & -
2 \left( \omega + \eta^2 \ajo{\omega}{\eta^2} \right)
 (\eta^0)' \Psi[C; \eta^0, \eta^1]
\label{eq23.2}
\eea

\noindent we can use the same procedure to obtain
 (\ref{eq1}) from (\ref{eq5}).

This concludes the proof of the equivalence. We can 
view (\ref{eq1}) and (\ref{eq5}) as
two equivalent wave functionals representing the exact
 quantum state 
$\mid \Psi_C>$.

\section{A third representation for the exact quantum 
states $\mid \Psi_C>$}

A third representation for the exact quantum states
 $\mid \Psi_C>$ can be obtained by
following a similar procedure as the one used in \cite{1}.

Notice that from the action (\ref{eq3}) the following
 secondary first-class constraints
can be obtained \cite{1}:

\bea
\rho'\Pi_{\rho} + \phi'\Pi_{\phi} - \Pi'_\rho \approx 0
\label{eq27} \\
(\phi')^2 - {\Pi^{2}_\rho \over 4} + (C - j(\phi))e^{2\rho}
 \approx 0
\label{eq28}
\eea

In the quantum theory, let us define the operators

\bea
\pihat_\phi & = & -i \hbar \ava{\;}{\phi} \\
\hat{\rho} & = & +i \hbar \ava{\;}{\Pi_\rho}
\eea

One can check that the wave functional,

\be
\Psi[C; \Pi_\rho , \phi] = \chi[C] \exp \left[ {i \over \hbar} 
\int_{x_-}^{x_+} dx
\left( \Pi_\rho + \phi' \ln \left( {2\phi' -
 \Pi_\rho \over 2\phi' + \Pi_\rho} \right)
- {\Pi_\rho \over 2} \ln \left( {(\phi')^2 - 
{\Pi^2_{\rho} \over 4} \over j(\phi) - C}
\right) \right) \right]
\label{eq31}
\ee

\noindent satisfies the quantum constraints:

\bea
\left( \Pi_\rho \hat{\rho}' + 
\phi' \hat{\Pi}_\phi - \Pi'_\rho \right)
\Psi[C; \Pi_\rho , \phi] & \equiv & 0
\label{eq32} \\
\left( \hat{\rho} -
\half \ln \left( { (\phi')^2  - 
{\Pi^2_{\rho} \over 4} \over
j(\phi) - C} \right) \right) 
\Psi[C; \Pi_\rho , \phi] & \equiv & 0
\label{eq33}
\eea
 
Equation (\ref{eq31}) is indeed another
 representation for the exact quantum states \\
$\mid \Psi_C>$. 

The equivalence of (\ref{eq31}) with (\ref{eq1})
 and (\ref{eq5}) can be established 
using the procedure  presented in section 2.

For completeness we report the generating functional
 $F[\Pi_\rho, \phi; \eta^0, \eta^1]$
of the canonical transformation (\ref{eq17.1} - \ref{eq17.4}):

\bea
F[\Pi_\rho, \phi; \eta^0, \eta^1] & = & 
\int_{x_-}^{x_+} dx \left[ {\Pi_\rho \over 2} \ln \left(
{e^2 \eta^2 \over \Pi^2_{\rho} - 
4(\phi')^2} \right) + \phi' \ln \left(
{2\phi' - \Pi_\rho \over 2\phi' + \Pi_\rho} \right) - 
\right. \none
& & \  \hspace{2.0cm}
- \left. {2 (\phi - \alpha) \over \eta^2}(\eta^0(\eta^1)' - 
\eta^1(\eta^0)') \right]
\label{eq35}
\eea
 
\noindent where $e$ is the Euler number.

\section{Conclusions}

We have proven that the quantum wave functionals in the
 CJZ variables of \cite{7} and
the wave functionals in geometric variables of \cite{1}
 are exactly equivalent. We also
presented a third representation of the exact quantum 
states for generic 2d dilaton
gravity $\mid \Psi_C>$ in terms of the dilaton field 
and the momentum conjugate to the
spatial component of the metric tensor.

Construction of the quantum theory also requires
 defining a scalar product on the 
subspace spanned by the physical states $\mid \Psi_C>$. 
I have not 
addressed this problem here.

The CJZ representation of the exact quantum states 
for 2d gravity coupled to an
abelian field can be obtained, using the approach
 presented here, directly from the
expression given in \cite{10}. 

\vspace{1.0cm}

\par\noindent
{\large\bf Acknowledgements}
\par
I am very grateful to Gabor Kunstatter for letting
 me know the results of his paper
prior to its publication and for useful conversations
 and comments. 
I would also like to thank Gordon Semenoff for useful
 conversations.
This work was supported by the National Sciences and 
Engineering Research Council of Canada.

\end{document}